\documentclass[twocolumn,showpacs,preprintnumbers,amsmath,amssymb]{revtex4}

\usepackage{graphicx}
\usepackage{dcolumn}
\usepackage{bm}

\begin{document}


\title{Observation of excited states in $^{20}$Mg sheds light on nuclear forces and shell evolution} 

\author {J.S. Randhawa$^{1,2}$, R. Kanungo$^{1,2}$,  M. Holl$^{1,2}$, J. D. Holt$^2$, P. Navratil$^2$, S. R. Stroberg$^{2}$, G. Hagen$^{3,4}$, G. R. Jansen$^{5,3}$, M. Alcorta$^{2}$, C. Andreoiu$^6$, C. Barnes$^1$, C. Burbadge$^7$, D. Burke$^8$, A. A. Chen$^8$, A. Chester$^6$, G. Christian$^2$, S. Cruz$^2$, B. Davids$^{2,9}$, J.Even$^2$\footnote{Present address: KVI-CART, University of Groningen, 9747 AA Groningen, The Netherlands}, G. Hackman$^2$, J. Henderson$^2$, S. Ishimoto$^{10}$, P. Jassal$^{1,2}$, S. Kaur$^{1,11}$, M. Keefe$^1$, D. Kisliuk$^7$, R. Kr\"ucken$^{2,{12}}$, J. Liang$^8$, J. Lighthall$^2$, E. McGee$^7$, J. Measures$^2$, M. Moukaddam$^2$, E. Padilla-Rodal$^{13}$, A. Shotter$^{14}$, I.J. Thompson$^{15}$, J. Turko$^7$, M.Williams$^{2,16}$, O. Workman$^1$}

\affiliation{$^1$Astronomy and Physics Department, Saint Mary's University, Halifax, NS B3H 3C3, Canada}
\affiliation{$^2$TRIUMF, Vancouver, BC V6T2A3, Canada}
\affiliation{$^{3}$Physics Division, Oak Ridge National Laboratory, Oak Ridge, Tennessee 37831, USA}
\affiliation{$^4$ Department of Physics and Astronomy, University of Tennessee, Knoxville, TN 37996, USA} 
\affiliation{$^5$ National Center for Computational Sciences, Oak Ridge National Laboratory, Oak Ridge, Tennessee 37831, USA}
\affiliation{$^6$Department of Chemistry, Simon Fraser University, Burnaby, BC V5A 1S6, Canada}
\affiliation{$^{7}$Department of Physics, University of Guelph, Guelph, ON N1G 2W1, Canada}
\affiliation{$^{8}$Department of Physics and Astronomy, McMaster University, Hamilton, ON L8S 4M1, Canada}
\affiliation{$^9$Department of Physics, Simon Fraser University, Burnaby, BC V5A 1S6, Canada}
\affiliation{$^{10}$High Energy Accelerator Research Organization (KEK), Ibaraki 305-0801, Japan}
\affiliation{$^{11}$Department of Physics and Atmospheric Science, Dalhousie University, Halifax, NS B3H 4R2, Canada}
\affiliation{$^{12}$Department of Physics and Astronomy, University of British Columbia, Vancouver, BC V6T 1Z1, Canada }
\affiliation{$^{13}$Instituto de Ciencias Nucleares, UNAM, AP 70-543, 04510 Mexico City, Mexico}
\affiliation{$^{14}$University of Edinburgh, Edinburgh, United Kingdom}
\affiliation{$^{15}$Lawrence Livermore National Laboratory, L-414, Livermore, California 94551, USA}
\affiliation{$^{16}$Department of Physics, University of York, Heslington, York, United Kingdom, YO10 5DD}

\date{\today}

\begin{abstract}
The exotic Borromean nucleus $^{20}$Mg with $N$ = 8, located at the proton drip-line 
provides a unique testing ground for nuclear forces and the evolution of shell structure in the neutron-deficient region. We report on the first observation of proton unbound resonances together with bound states in $^{20}$Mg from the $^{20}$Mg($d$,$d'$) reaction performed at TRIUMF. Phenomenological 
shell-model calculations offer a reasonable description. However, our experimental results present a challenge for current
first-principles nuclear structure approaches and point to the need for improved chiral forces and {\it ab initio} calculations. 
Furthermore, the differential cross section of the first excited state is compared with distorted-wave Born approximation calculations to deduce a neutron quadrupole deformation parameter of $\beta_n$=0.46$\pm$0.21.
This provides the first indication of a possible weakening of the $N$ = 8 shell closure at the proton drip-line. 

\end{abstract}

\pacs{24.50+g, 25.45.De,  25.60.-t, 25.70.Ef }
\maketitle

The evolution of shell structure over the nuclear landscape is a manifestation of strong interactions in the complex nuclear many-body system. Properties of nuclei at the neutron and proton drip-lines provide new arenas to investigate the effects of large proton-neutron asymmetry and understand the persistence of mirror symmetry. Shell structure evolution in neutron-rich and proton-rich nuclei \cite{JA05,JA09,OZ00,HO08,KA09, PI01,NA00,MO95,BA07,WIE13,ST13,GAR16} are leading to new insights into nuclear forces, including the role of three-body forces \cite{OT10,HEB15}. The region around the $N$ = 8 shell closure draws particular interest, since this shell gap disappears at the neutron drip-line and leads to the formation of a two-neutron halo in the Borromean nucleus $^{11}$Li. The small two-neutron separation energy ($S_{2n}$ = 360 keV) of  $^{11}$Li results in the excited states of $^{11}$Li being unbound. Less is known about the structure of nuclei at the proton drip-line. The $N$ = 8 isotone at the proton drip-line, $^{20}$Mg, is also a Borromean system whose two-proton separation energy is $S_{2p}$ = 2.337(27) MeV \cite{AME12}. There is no experimental information on resonances above the proton threshold in $^{20}$Mg. In this work we present the first observation of a resonance in $^{20}$Mg through deuteron inelastic scattering. This measurement provides new insight into shell evolution as well as tests of {\it ab initio} predictions. The resonance(s) in $^{20}$Mg could also contribute to a potential breakout path from the hot CNO cycle via two-proton capture on the waiting point nucleus $^{18}$Ne in Type-I x-ray bursts \cite{GO95}. 

Microscopic cluster model \cite{DE98} predictions for the  2$^+$ state agree with the experiment \cite{GA07,IW08}. A 4$^+$ state is predicted  at 3\,MeV, this state is predicted around 3.8 MeV, using a beyond the mean field approach \cite{RO16}. Predictions based on an angular momentum projected generator coordinate framework \cite{RO02},  find the 2$_1^+$ and the  4$_1^+$ states to be higher in energy around, 3.5 MeV and 7.8 MeV, respectively.  
{\it Ab initio} calculations in a many-body perturbation theory (MBPT) framework including the three-nucleon (3N) force \cite{HO13} predict the first excited state in fair agreement with Refs.\cite{DE98,RO16}. The predictions widely vary however, for states above the proton threshold with the $2_2^+$ and 4$_1^+$ states being nearly degenerate placed around 4.2 MeV. 
Experimental information on these states above the proton threshold, is therefore necessary for testing the nuclear structure models and nuclear forces. 

In addition, mass measurements of $^{20,21}$Mg \cite{GA14} cannot be reconciled with the known isobaric mass multiplet equation for $^{20}$Mg from isospin symmetric shell-model Hamiltonians \cite{BA06} or with that including isospin non-conserving interactions \cite{OR89}. The {\it ab initio} predictions based on MBPT \cite{HO13} show a stronger isospin dependence of the IMME than is experimentally observed \cite{GA14}. The energy of the lowest $T$ = 2, $J^\pi$ = $0^+$ state in $^{20}$Na measured from superallowed $\beta^+$ decay of $^{20}$Mg however successfully describes the $T$ = 2, mass 20 multiplet by a quadratic IMME  \cite{GL15}. Excited states of the mirror nuclei $^{20}$Mg and $^{20}$O can provide further insight into the isospin dependence of the nuclear interaction and the nature of the $N$ = 8 shell closure at the proton drip-line.

The first excited state of $^{20}$Mg is a bound state that was observed first through gamma ray detection from the $^9$Be($^{22}$Mg,$^{20}$Mg+$\gamma$)X reaction \cite{GA07} exhibiting a peak at 1.598(10) MeV.  A gamma transition at 1.61(6) MeV was also observed in the Coulomb excitation of $^{20}$Mg with a Pb target at 58.4$A$ MeV \cite{IW08}.  Assuming this as the 2$^+$ state, a B(E2; 0$^+ \rightarrow$ 2$^+$) value of 177(32) e$^2$fm$^4$ was deduced \cite{IW08}, which is in good agreement with predictions in Refs.\cite{RO02} as well as the cluster model predictions \cite{DE98}. In order to obtain insight on the deformation of neutrons in this $N$ = 8 isotone, inelastic scattering with a hadronic probe is needed together with information derived from Coulomb excitation. 

In this article, we report the first study of deuteron inelastic scattering on $^{20}$Mg, populating new resonances together with the ground and first excited states in this proton drip-line nucleus. The experiment was carried out using the IRIS reaction spectroscopy facility \cite{KA13} at TRIUMF. A schematic view of the detector layout is shown in Fig.1. The radioactive beam of $^{20}$Mg was produced via fragmentation of a SiC target with 480 MeV protons. The beam was re-accelerated using the ISAC-II superconducting LINAC \cite{Bob} to 8.5$A$ MeV and then passed through an ionization chamber, filled with isobutane gas at 19.5 Torr at room temperature. The energy loss of the beam, measured in this detector, provided an event by event identification of the $^{20}$Mg incident beam and its contaminant $^{20}$Na throughout the experiment. Following this, the beam interacted with a thin windowless solid deuterium $^{2}$H ($D_2$) reaction target built on a 4.5 $\mu$m thick Ag foil backing facing upstream of the $D_2$ layer. The target cell with the foil was cooled to 4 K, before forming solid $D_2$. The energy of the elastic scattered beam on the Ag foil was measured with and without $D_2$, providing a continuous measurement of the target thickness during the experiment. These scattered beam particles were detected using a double sided silicon strip detector placed 33 cm downstream of the target, covering laboratory angles  1.9$^{\circ}$ to 6.1$^{\circ}$. The average target thickness was  65 $\mu$m,  and the value at each instant of time (each data collection run) was used for determining the scattering cross sections. The deuterons scattered out from reactions were detected using annular arrays of 100 $\mu$m thick single sided silicon strip detectors followed by a layer of 12 mm thick CsI(Tl) detectors.  This detector combination served as an energy-loss ($\Delta$E) and total energy (E) telescope for identifying the $p, d, t$ and He recoils after the target. The CsI(Tl) detectors were calibrated using the elastic and inelastic scattering from a beam of $^{20}$Ne. The detector telescope covered scattering angles of $\theta_{lab}$ = 30.1$^{\circ}$ to 56.2$^{\circ}$. 

\begin{figure}
\includegraphics[width=8cm, height=4cm]{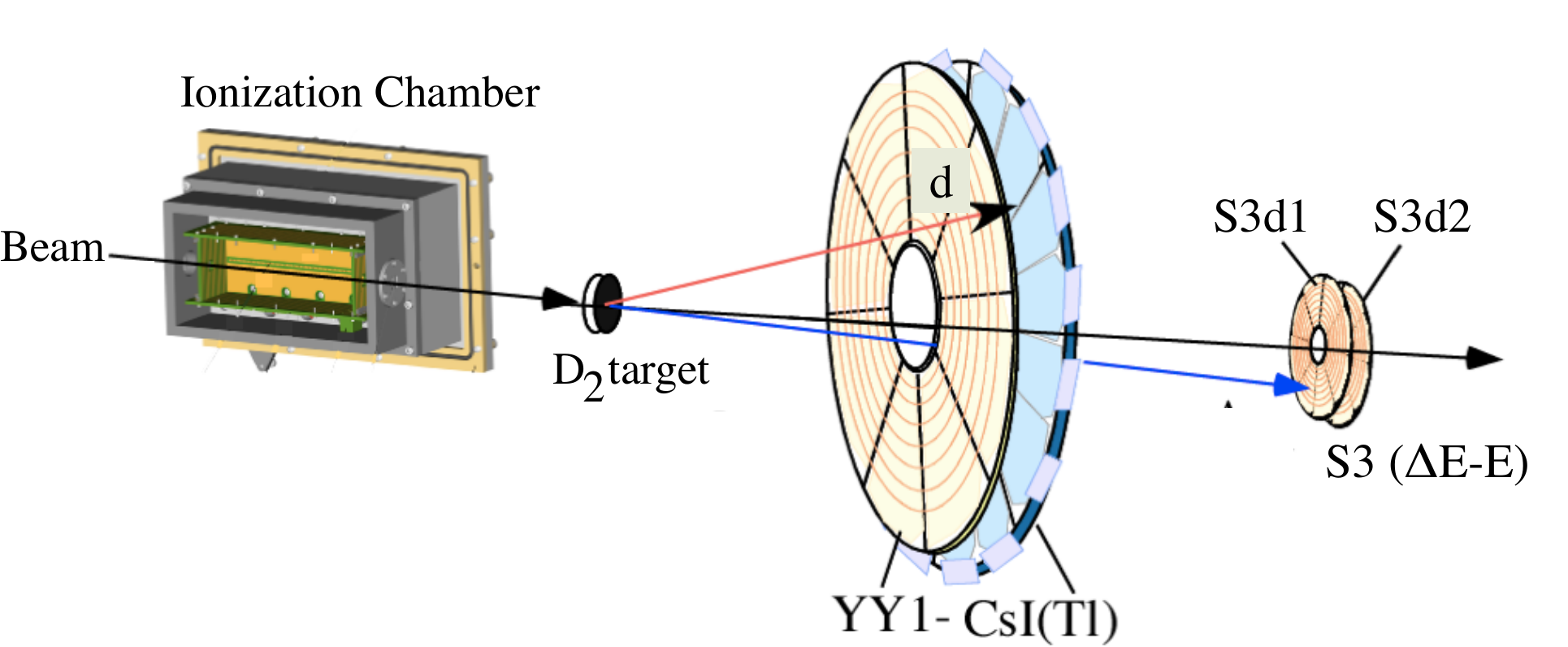}
\caption{\label{fig:epsart}  Schematic view of experimental setup at IRIS. }
\end{figure}

The excitation spectrum of $^{20}$Mg, shown in Fig.2a,  was reconstructed using the missing mass technique from the energy and scattering angle of the deuterons, measured by the silicon-CsI(Tl) telescope. The deuteron scattering can detect proton unbound resonances with no background from decay protons, unlike proton inelastic scattering. The ground state peak from $^{20}$Mg($d$,$d$) elastic scattering is clearly visible together with two prominent peaks, one below and one above the proton threshold. The background from the Ag foil was measured by collecting data without $D_2$ and is shown with the green dashed-dotted histogram normalized by the incident beam intensity. This background contribution was $\approx$ 10\% of the total spectrum. The non-resonant background from $^{20}$Mg+$d\rightarrow^{18}$Ne+$p$+$p$+$d$ estimated from Monte Carlo simulation, including the detection conditions, is shown by the blue short dashed histogram. The total background from these two contributions is depicted by the red dashed histogram and is normalized to the data in the excitation energy region greater than 6.5 MeV. 

\begin{figure}
\includegraphics[width=7cm, height=9cm]{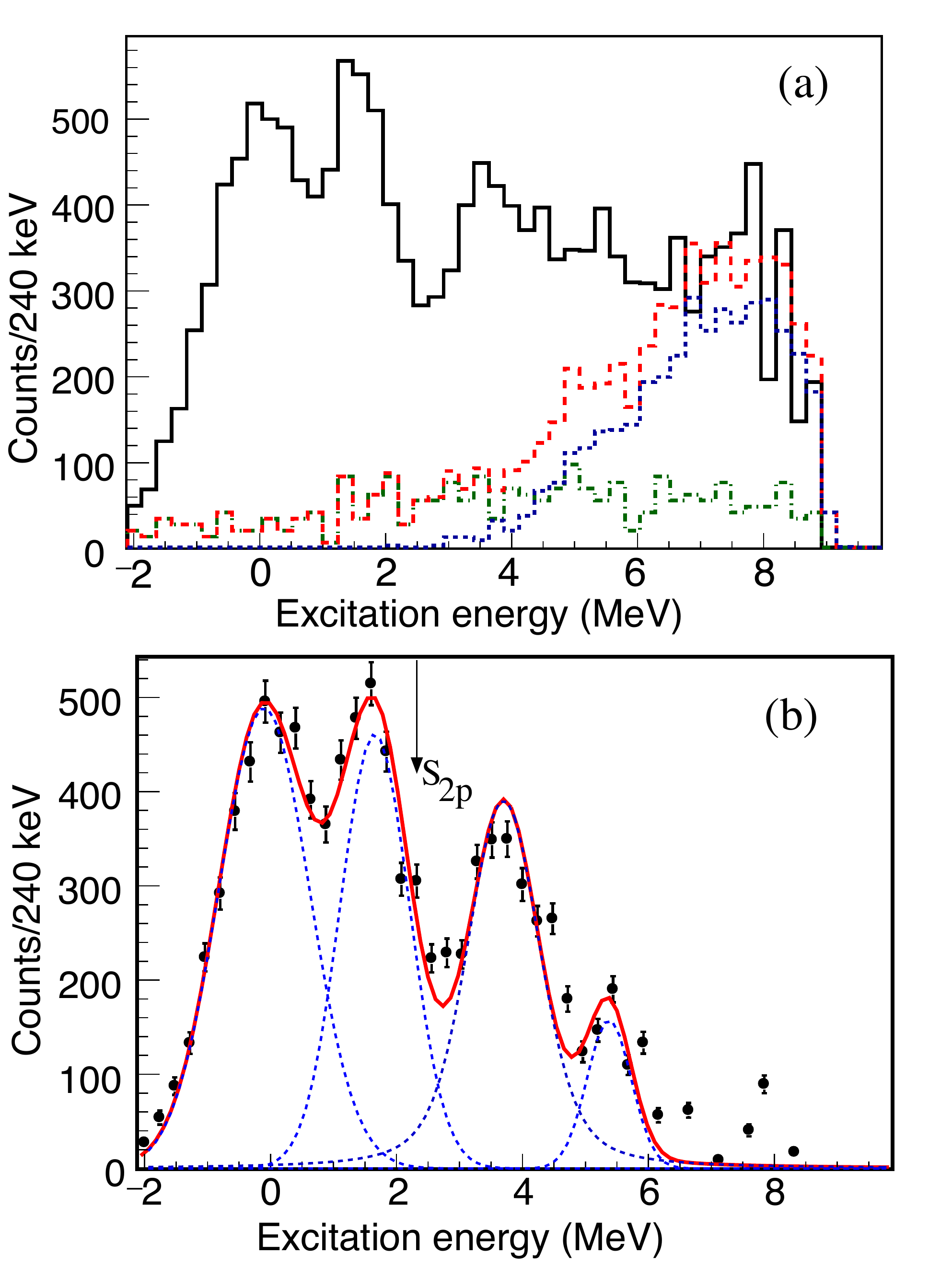}
\caption{\label{fig:epsart}  (a) The excitation energy spectrum for $^{20}$Mg measured from $^{20}$Mg($d$,$d'$) reaction (black histogram). The measured background from Ag backing foil (green dashed-dotted histogram). Background from $^{20}$Mg+$d\rightarrow$$^{18}$Ne+$p$+$p$+$d$ four-body phase space (blue dashed histogram). The red histogram shows the total background (sum of blue and green histograms). (b) Background subtracted excitation energy spectrum for $^{20}$Mg. The curves show the results of fitting (see text). The blue dashed curves are the individual fitting components and the red curve is the their sum.}
\end{figure}

The background subtracted excitation spectrum (Fig.2b), was fitted by a sum of two Gaussians, for the two bound state peaks, and two Breit-Wigner functions folded by Gaussian profiles accounting for the detection resolution, for the unbound resonance peaks. The excitation energy resolution ($\sigma$) for the ground state peak was  0.71 MeV  for $\theta_{lab}$= 35$^{\circ}$ and 0.45 MeV for $\theta_{lab}$= 50$^{\circ}$, which is in fair agreement with simulations. The resolution improves for the excited states to  0.48 MeV for the newly observed second excited state as determined based on simulations. 
The excitation energy of the first excited state peak is found to be  1.65$^{+0.02}_{-0.10}$ MeV which is in good agreement with the observations in Refs.\cite{IW08, GA07}. 
Above the proton threshold, the most prominent peak is observed at an excitation energy of  3.70$^{+0.02}_{-0.20}$ MeV with an  
apparent resonance width of 0.47$\pm$0.06 MeV after unfolding the resolution. 
The excitation energy uncertainties include observable peak shifts for possible systematic effects besides peak fitting. 
A small structure is also observed at a higher excitation energy of 5.37$\pm$0.02 MeV where the uncertainty quoted is from fitting only.

The energies of the first two excited states states observed in $^{20}$Mg are in close agreement with those of the mirror nucleus $^{20}$O \cite{LA73}. Shell model calculations with the phenomenological USDB \cite{BA06} interaction without any isospin dependence are in fair agreement with the data (see Fig.~3), though the first excited state is predicted slightly higher in energy and the second excited state observed seems closer to the predicted 4$^+$ state. Calculations with an improved USDB interaction, including the Coulomb corrections and isospin dependence of Ref.~\cite{HET89}, provide excellent agreement with the first excited (2$^+$) state while the observed second excited state is midway between the predicted 4$^+_1$ and 2$_2^+$ states. Given the small predicted energy difference of these states, it may be possible that the new resonance peak observed in the experiment is an overlap of these two states. Furthermore, valence-space calculations within the MBPT framework, based on NN+3N forces \cite{HO13} (within an extended proton valence space that included the $sd$ shell plus $f_{7/2}$ and $p_{3/2}$ orbitals), also predict the first excited $2^+$ energy in good agreement with experiment, but higher-lying states are several hundred keV above the observed resonances. This could be expected since these calculations neglect effects of continuum coupling, which typically lower states by a few hundred keV.

\begin{figure}
\includegraphics[width=8cm, height=8cm]{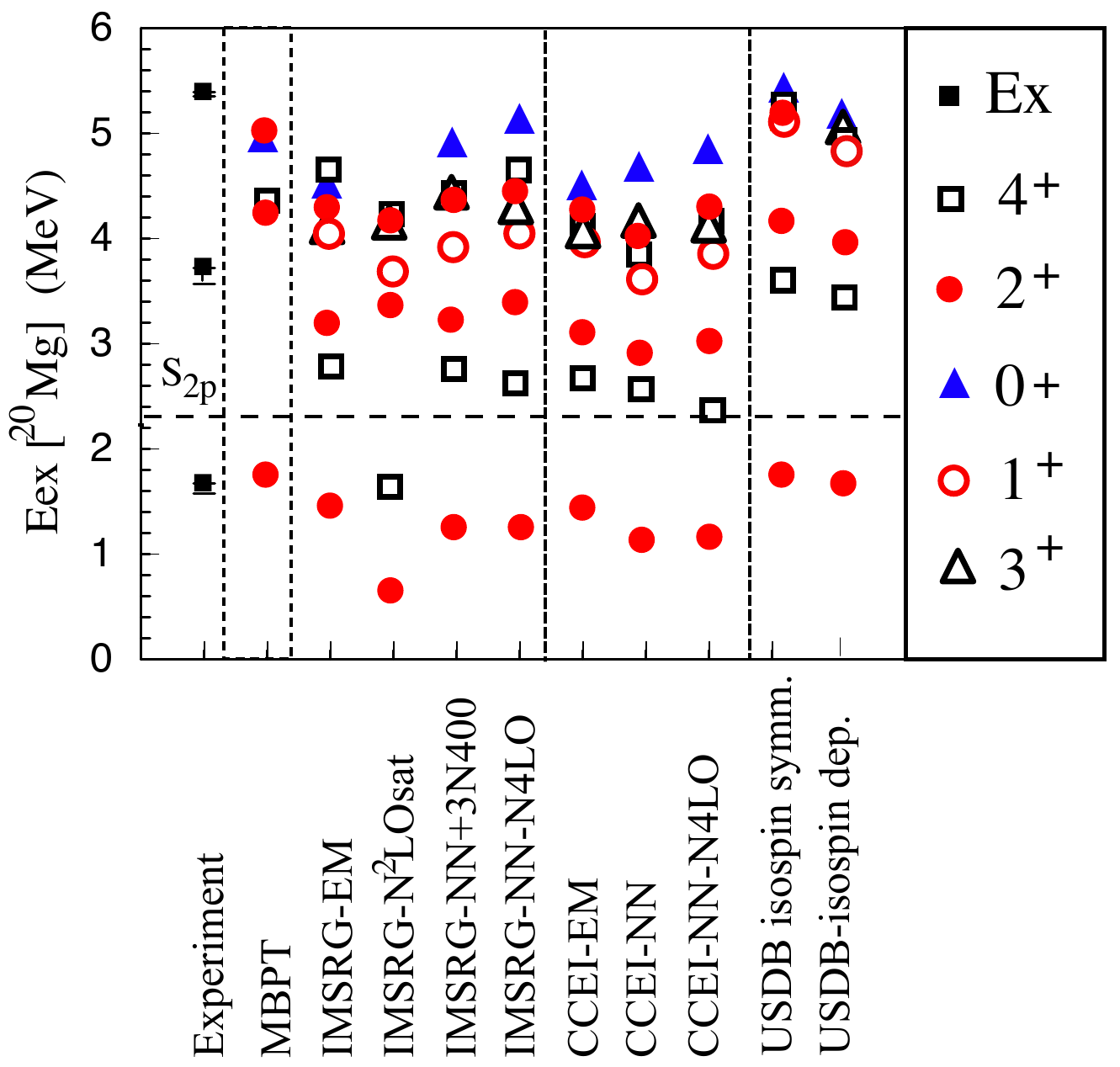}
\caption{\label{fig:epsart}  Measured excited states (this work) in $^{20}$Mg (black filled squares) compared to theoretical predictions.  {\it Ab initio} predictions in MBPT framework with the chiral NN+3N interaction \cite{HO13}, IMSRG framework with EM, N$^2$LO$_{\rm{sat}}$, NN+3N(400) and NN-N$^4$LO+3N(lnl) interactions,  CCEI framework with the EM, NN+3N(400) and NN-N$^4$LO+3N(lnl)  interactions, and shell model with USDB interactions without and with isospin dependence. The spins of the predicted states are indexed in the figure.}
\end{figure}

New {\it ab initio} calculations were therefore performed. In particular, we use the valence-space in medium similarity renormalization group (VS-IMSRG) approach \cite{TSU12,BOG14,STR16}, and the coupled-cluster effective interaction (CCEI) method \cite{JA14,JA16}. The VS-IMSRG approach constructs an approximate unitary transformation \cite{MOR15,HER16} to first decouple the $^{16}$O core, as well as an $sd$ valence-space Hamiltonian, diagonalized using the NuShellX@MSU shell-model code \cite{BRO14}. We further capture the bulk effects of 3N forces between valence nucleons with the ensemble normal-ordering procedure of Ref.~\cite{STR17}, thereby producing a unique valence-space Hamiltonian for each nucleus to be studied. This allows us to test nuclear forces in essentially any fully open-shell system accessible to the nuclear shell model with a level of accuracy comparable to large-space {\it ab initio} methods. In the CCEI approach, similar to the VS-IMSRG method, one calculates the valence-space effective interaction starting from a chiral NN+3N interaction and applies the obtained zero-plus-one-plus-two-body interaction in the standard shell model diagonalization. The CCEI approach~\cite{JA16} utilizes coupled-cluster method to perform {\it ab initio} calculations for $A_c$, $A_c+1$ and $A_c+2$ system with $A_c$ the number of nucleons in the core (here $A_c=16$). 
The Okubo-Lee-Suzuki similarity transformation \cite{OK54,SU82} is then used to obtain an effective $sd$-shell interaction.

\begin{figure}
\includegraphics[width=8cm, height=11cm]{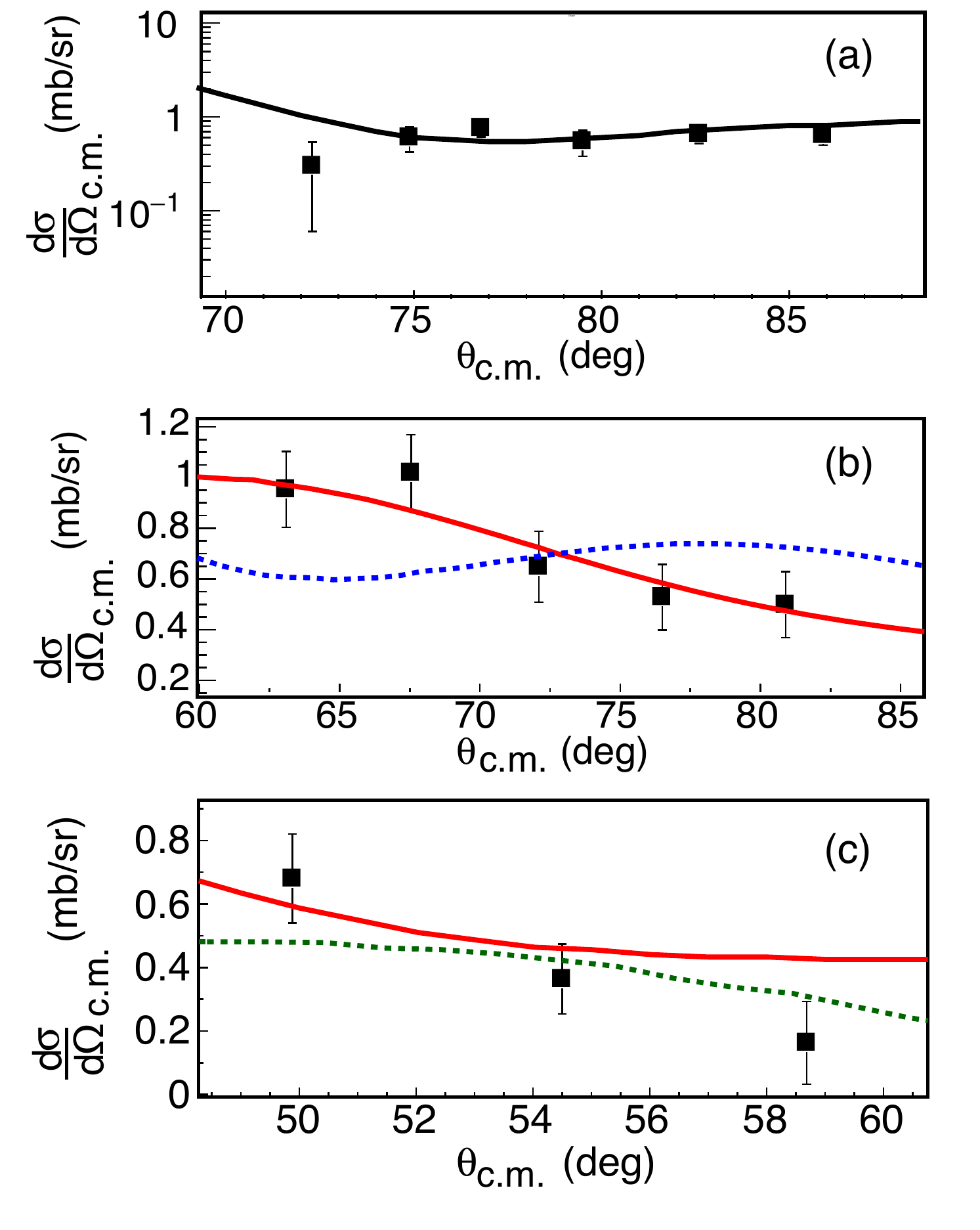}
\caption{\label{fig:epsart}  Angular distribution data (black filled circles) in the center-of-mass frame (c.m.) for (a) $^{20}$Mg$_{gs}$ (b) $^{20}$Mg$_{ex}$; 1.65 MeV (c) b) $^{20}$Mg$_{ex}$; 3.7 MeV. The black curve in (a) is a calculation using phenomenological optical model potential. The DWBA calculations are shown by red solid / blue dotted curves  for $L$ = 2 / $L$ = 3 in (b) and  red solid/green dotted curve for $L$ = 2 / $L$ = 4 in (c).}
\end{figure}

We take several sets of initial NN+3N forces in this work. The first, EM(1.8/2.0) \cite{HEB11,SIM16}, begins from the chiral NN N$^3$LO force \cite{EN03}  with a nonlocal 3N force fit in $A$ = 3 and 4-body systems, but reproduces ground- and excited-state energies to the tin region and beyond \cite{SIM17,MOR18}. The second, NN+3N(400) begins from the same NN force, but with local 3N forces \cite{NA07,RO12}, yields accurate binding energies and spectra in and around the oxygen isotopes \cite{BOG14,JA14,HE13,CI13}. Next, we utilized a newly developed chiral potential at N\textsuperscript{4}LO~\cite{PhysRevC.96.024004} combined with an N$^2$LO 3N interaction with parameters fitted to the $^3$H binding energy and half-life (NN-N$^4$LO+3N(lnl)). Finally N$^2$LO$_{\rm{sat}}$ has been fit to medium-mass data and reproduces ground-state energies and radii to the nickel region \cite{EKS15}. 

VS-IMSRG calculations based on the EM(1.8/2.0) interaction accurately predicts the energy of the first $2^+$ state, but states above the proton threshold are lower than the experimental observations as well as the shell model predictions. The N$^2$LO$_{\rm{sat}}$ interaction has the worst agreement predicting the first excited state at only 600 keV. The other two interactions, NN+3N400 and NN N$^4$LO+3N(lnl),  provide similar results as EM(1.8/2.0), but with slightly lower 2$^+_1$ energies and a greater spread in higher-lying states.

The MBPT results use older generation NN+3N forces and an extended valence space. We therefore compare the VS-IMSRG and CCEI frameworks where very similar results are obtained with the same forces, with the VS-IMSRG spectrum being modestly more spread. This is likely due to the inclusion of repulsive 3N forces between the four valence protons via the ensemble normal ordering procedure in the VS-IMSRG approach. There is a greater variance in the predicted energies for different nuclear forces.  

We note here that, binding energy from the CCEI calculations with NN+3N(400) for $^{20}$Mg agree with the experiment.
To further check the VS-IMSRG and CCEI calculations, we also performed no-core shell model (NCSM) \cite{BA13} calculations with the same interactions. For technical reasons, the largest basis space we could reach was $N_{\rm max}=6$ (utilizing the importance-truncation approach \cite{RO07}). Such a space is too small to reach convergence and obtain a reliable estimate of excitation energies. However, our NCSM results within their uncertainties were consistent with the valence-space methods for binding energy as well as for the $2^+_1$ state excitation energy. 

The observed unbound state of $^{20}$Mg therefore, points to the need for refinement of the nuclear forces although the IMSRG framework with the EM interaction (3N-full) in Ref.\cite{STR16} was shown to explain the spectra for Ne and F isotopes. While, the role of continuum effects needs to be assessed it should be noted that coupling to the continuum generally lowers the excitation energy as shown in Ref.\cite{HA16}. Furthermore, the shell model predictions without any coupling to the continuum agree well with the data.   

The differential cross sections of the ground state, first and second excited states were derived from the area under the background subtracted peaks for different angular bins (Fig.4). The Woods-Saxon shape optical potential parameters for the $^{20}$Mg+$d$ interaction were determined from the best fit of DWBA calculations to the elastic scattering angular distribution data (Fig.4a) using sFRESCO \cite{TH}. These parameters were then used for calculating the angular distributions for the excited states. The angular distribution of the first excited state (Fig.4b) is consistent with a multipolarity of excitation $L$ = 2, thereby determining for the first time its spin of 2$^+$. The normalization of the calculation to the data provides the deformation length $\delta$ to be 1.33$\pm$0.23 fm, where $\delta = {\frac{Z}{A}}\delta_p + {\frac{N}{A}}\delta_n$ with $\delta_p$ and $\delta_n$ being the proton and neutron deformation lengths, respectively. The deformation parameter is given by $\beta$ = $\delta$/$R$ with $R$ being the radius. $\delta_p$ was determined to be 1.32$\pm$0.12 fm from the quadrupole proton deformation parameter 0.44(4) derived from Coulomb excitation \cite{IW08} and a proton radius value of 3 fm that is consistent with theoretical predictions \cite{Gai14,WA07,LA01}. Therefore, a quadrupole deformation parameter for neutrons, $\beta_{n}$ = 0.46$\pm$0.21 is found. This large non-zero value within one standard deviation uncertainty, depicts neutron deformation and therefore a first signature of possible weakening of the $N$ = 8 shell closure. The  $\beta_{n}$ for $^{20}$Mg is in agreement with $\beta$  of the mirror nucleus $^{20}$O found from proton scattering \cite{JE99}. Using the prescription of \cite{JE99} $\beta_{n}$ = 0.46$\pm$0.14 for $^{20}$Mg. The proton deformation parameter of $^{20}$O obtained from B(E2) measurements \cite{AD16} is also in agreement. 

The angular distribution for the second excited state (Fig.4c), is not explained by either $L$ = 2 ($\chi^2_{red}\approx$ 2.7) or $L$ = 4  ($\chi^2_{red}\approx$ 4.1) excitation. There may be a possibility that the resonance peak has a mixture of $L$ = 2 and 4.  The observed resonance energy lies between the predicted 2$_2^+$ and 4$^+$ states with the isospin dependent USDB interaction reflecting the importance of isospin dependence. 

In summary, the first observation of proton unbound excited states in $^{20}$Mg at 3.70$^{+0.02}_{-0.20}$ MeV and around 5.37 MeV from deuteron inelastic scattering is reported. The new data present a challenge for different nuclear forces from chiral effective field theory and {\it ab initio} calculations in the VS-IMSRG and CCEI frameworks since the results underpredict the observed resonance energy, although ground state binding energies were explained. This first systematic study shows that some of the different force prescriptions exhibit larger variance in the predicted energy, while the predictions from the two many body methods are similar using the same starting forces.
The spin of the first excited state is determined to be 2$^+$ and a neutron quadrupole deformation parameter  $\beta_{n}$ = 0.46$\pm$0.21 is found from its differential cross section. This large deformation provides first signature of possible weakening of the $N$ = 8 shell closure at the proton drip-line. 
The data will therefore motivate further experimental and theoretical studies to explain the observed features.

The authors express sincere thanks to the TRIUMF beam delivery team. The support from NSERC, Canada Foundation for Innovation and the Nova Scotia Research and Innovation Trust is gratefully acknowledged. TRIUMF receives funding via a contribution through the National Research Council Canada. The support from RCNP for the target is gratefully acknowledged. It was partly supported by the grant-in-aid program of the Japanese government under the contract number 23224008 and 14J03935. 
We thank J. Simonis, K. Hebeler, and A. Schwenk
for providing the EM interaction 3N matrix elements used in this work
and for valuable discussions.  Computations were performed 
with an allocation of computing resources at the
J\"ulich Supercomputing Center (JURECA). This work was performed under the auspices of the U.S. Department of Energy by Lawrence Livermore National Laboratory under Contract DE-AC52-07NA27344. This material is based upon work supported by the
U.S. Department of Energy, Office of Science, Office
of Nuclear Physics under Award Number DE-SC0018223 (SciDAC-4 NUCLEI) and the Field Work Proposals ERKBP57
and ERKBP72 at Oak Ridge National Laboratory
(ORNL). This research used resources
of the Oak Ridge Leadership Computing Facility located
at ORNL, which is supported by the Office of Science
of the Department of Energy under Contract No. DE-
AC05-00OR22725. J.E. gratefully acknowledges financial support from the German Academic Exchange Service (DAAD Postdoc program).


\end{document}